# COLLABORATIVE BIBLIOGRAPHIC SYSTEM FOR REVIEW/SURVEY ARTICLES

Mutaz Beraka, Abdullah Al-Dhelaan, Mznah Al-Rodhaan

Department of Computer Science
College of Computer & Information Sciences
King Saud University, Riyadh, Saudi Arabia

***ABSTRACT***

*This paper proposes a Bibliographic system intends to exchange bibliographic information of survey/review articles by relying on Web service technology. It allows researchers and university students to interact with system via single service using platform-independent standard named Web service to add, search and retrieve bibliographic information of review articles in various science and technology fields and build-up a dedicated database for these articles in each science and technology field. Additionally, different implementation scenarios of the proposed system are presented and described, andrich features that offered by such system are studied and described. However, this paper explains the proposed system using computing area due to the existence of detailed taxonomy of this area, which allows defining the system, their functionalities and features provided.However, the proposed system is not only confined to computing area, it can support any other science and technology area without any need to modify this system.*

## 1. INTRODUCTION

The revolution in scientific research around the world produced a tremendous amount of scientific research papers that published daily by researchers and students in a variety of journals and conference proceedings. Survey or review article is one type of scientific paper that summarizes the state of the art on specific topic or field. The value of this article is massive especially if such paper published in reputation journal or conference. It gives the reader fast-pass card to enter specific topic and summarize the findings of published work to highlight advances and new research lines. In the current decade, there are a good number of review articles that were published and were indexed in different indexing services. Searching about this kind of papers are not an easy job and time consuming. It requires extensive searching process in different digital libraries, search engines, journals and others to find what you are looking for, especially in the presence of a huge number of different topics and their sub-topics. This process is a premonition for many researchers and students due to it is a compulsory step to start thinking about new ideas for any scientific research paper, project or thesis proposal. Therefore, the biggest question that going on our minds is how we can make this process easier as possible in the absence of database or search engine dedicated to bibliographic information of scientific review articles? The best answer is by building and developing one bibliographic system that allows researchers and students to add their bibliographic information of review articles in various science and technology subject area using service-orientation paradigm and Web service technology and building-up the database dedicated for scientific review articles of each field/area and feeding them at the same time. This allows users to search exclusively about review articles in specific science and technology field. In [1], we briefly presented the proposed system and its features whereas in this paper, we extensively describe the proposed system, its different implementation scenarios and its features. In addition, we suggest some future improvements for our proposed system.

Service Oriented Architecture (SOA) as a fundamental design-principle allows providing and exposing application in form of interoperable services available in a network such as the World Wide Web [2]. Implementation of SOA can involve designing and developing applications that reuse loose coupled services, applications available as services for different purposes [3]. However, SOA alone is a design principle and the most sophisticated and popular platform-independent standard technology that uses to implement a set of principles and methodologies found in SOA is a Web service technology. Web service is a software system designed to support interoperable machine-to-machine interaction over a network [2]. Moreover, these applications are compatible with different programming languages, operating systems, hardware platforms, and are accessible from any geographic location [4]. Web service is describing using Web Service Description Language (WSDL) and invoking using Simple Object Access Protocol (SOAP) messages that are commonly transmitting using HTTP protocol.

Based on the best of our knowledge, there is no database exists and exclusively dedicates for scientific review articles as well as no bibliographic system dedicates to search and retrieve bibliographic information of scientific review articles in different science fields. Thus, this paper proposes a Bibliographic system that uses Web service technology to allow researchers and students to add, search and retrieve bibliographic information of survey/review articles over Intranet/Internet. This system builds-up a dedicated database for bibliographies of scientific review articles for each science field using its classification system and feeds it through user's information that will be used later on by search engine.

# 2. PROBLEM STATEMENT

The evolution in information technology arena and the growth in research community increased the demand for designing, developing and implementing different bibliographic applications and systems that achieved various requirements to deal with bibliographic information. These systems vary from bibliographic information systems, bibliographic management systems, bibliographic retrieval systems, bibliographic classification systems, bibliographic visualization systems and others. Our focus in this research is bibliographic management systems that allow exchanging bibliographic metadata among scientists, researchers and students that located around the world. Several systems have been developed to tackle such matter, where the requirements from those systems are different, which led to different strategies of implementation and different capabilities of each system Examples of those systems are Bibster [5], Mendeley [6], ShaRef [7] andSOCIOBIBLOG [8].

These systems are by some means sharing a number of functionalities with the proposed system, but the proposed system is mostly intended to exchanging bibliographic information of review/survey scientific articles. The simplicity of the design of the proposed system:(1) makes it applicable to be deployed in different environments. (2) makes users easily pushing their bibliographic information in order to exchange it with others users as well as pulling bibliographic information of review articles that found in the system, (3) allows to retrieve and add survey/review articles from digital libraries such as IEEE Xplore, to system database,and (4) allows to use available bibliographic system Application Programming Interface API(s) such as Mendeley API to retrieve review articles only to add them into appropriate database. In addition, the proposed system relies on the classification system for each science and technology field, (not custom classification) that uses to categorize added articles according to their disciplines under this field. It also provides Bibliometrics information about each sub-field in particular area, rating feature,evaluation & feedback feature and recommendation feature as four additional features compared with other systems.However, as we known a review or a survey article attempts to summarize the current state of understanding on a topic. In other words, it summarizes the state-of-the-art on a specific topic. These papers are needed for researchers as well as for students to aware of what is going on in the field and what has done before, and serve as basis to derive a new idea or invention [9]. However, searching about review articles is time consuming and needs from researchers or students to access different digital libraries and search engines to find what they want. This extensive process triggers the need for developing a system that intends to search about review articles or search engine and retrieves only review or survey papers that match given keyword. To achieve that, a database that contains all bibliographic information and meta-data about such papers is needed in order to allow developing a searching system or extend the capability of existing search engine to search in this database. Based on the best of our knowledge, there is no database exists and exclusively dedicates for scientific review articles as well as no bibliographic system dedicates to search and retrievebibliographic information of scientific review articles in different science fields. Thus, this paper propose a Bibliographic system that uses Web service technology to allow researchers and students to add, search and retrieve bibliographic information of review articles over Intranet/Internet. This system builds-up a dedicated database for bibliographies of scientific review articles for each science field using its classification system and feeds it through user's information that will be used later on by search engine. Of course, each user of the system will be participated with his or her list of review papers to feed different databases with this information in case of these papers belong to different science fields. Each paper will be added under pre-defined field according to taxonomy of paper field after getting approval from administrator of the system. This procedure is required in case of the proposed system deployed in public environment (see Section 3) in order to prevent any attempt to add non-review or survey article or to add review article under wrong field.

Based on what we have mentioned before, the proposed system is very useful for any researchers and students in any science and technology subject area that allows them to retrieve only survey and review articles easily by facilitating the process of searching for those papers in specific field of interest. It will be more than useful for postgraduate students due to simplifying the process of searching about review articles in interested field (such as computing field), which is a compulsory step for them to enter such field in order to pick-up a new idea for thesis proposal later on. On other hands, the future vision of the proposed system for each supported science and technology subject area is to provide a dedicated bibliographic database for review articles of this area that will serve its community. The dedicated database for each area can serve as a base to develop a new specific search engine (in our case let us called it "search engine for computing review articles") or may be used by existing search engine to support a new kind of search that pertains only for review articles in particular area by extended search engine capabilities.

However, the proposed system aims to:

- Build cost-effective system for exchanging bibliographic information of review/survey articles in various science and technology subject areas.
- Build cost-effective database for each science and technology subject area that stores bibliographic information of review articles in this field.
- Help users to benefit from bibliographic information of review articlesofeach other's.
- Support any science and technology subject area by using its classification system (or subject classification) to list major fields and their sub-fields in order to facilitate the process of adding and searching review articles under/in this area. The absence of the classification system of any science and technology subject area prevents to supportsuch area by the proposed system. This may be considered as negative point at the first look, but by looking to it as positive point,, this will motivate the community of such area to develop and define such classification for that area, which will be considered as a contribution in that area.

- Provide Graphical User Interface (GUI) to ease-to-use a system
- Provide a set of rich features including bibliographic analysis, a rating feature, recommendation feature and evaluation & feedback feature.
- Promote the goal of separating service consumers from the service provider (service implementation), where service can be run on various distributed platform and be accessible across Internet.
- Provide high level of availability to functionalities of system through WSDL of Web service all the time.
- Builds a basic database that will be served as an initial database intends tobibliographic information of scientific review papers.
- Provide search engine for scientific review articles for non-system users beside the proposed system.
- Reuse database(s) for future improvements.

# 3. THE PROPOSED SYSTEM

The simplicity of proposed system encourages and invitesdifferent levels of researchers and students for adding, searching and retrieving bibliographic information of review articles to builds-up a dedicated database for each supported science and technology subject area such as computing area, and feeds those databases. A dedicated database for each area will be a first attempt to create particular database intends to bibliographic information of survey and review articles for such area; which maximizes beneficially among researchers in that community. For instance, creating a dedicated database for computing survey and review articles will maximizes beneficial among researchers and students in computing community. The proposed system can be implemented and deployed in two different environments, which are private/closed environment and public/opened environment with different implementation scenarios. These different options and scenarios of implementation are offered to cope with different demands.In private environment, such as a college in university, the proposed system can be implemented and deployed in four different scenarios. These implementation scenarios are described in the following paragraphs.

***Implementation Scenario 1:***the proposed system will be a private bibliographic system in private environment such as a college in university that allows students, faculty members and researchers within a college exchanging and retrieving bibliographic information of survey and review articles. All users of such system are assumed to be trusted in term of they will only publish their bibliographic information of survey and review articles they have, and none of them will add erroneous information. In that case, the roles of administrator or let us named it as moderator will be confined only in adding and updating the classification system(s) of science and technology area(s), and it will not be responsible for approving or rejecting the bibliographic information of review articles added by users; which will become the responsibility of user's themselves. Thus, the moderator application could not be needed in this environment, and one of system users aka "Associate user" will take the role of adding and updating the classification system(s) of science and technology area(s) along with its normal roles. The architecture of the proposed system (Scenario 1) is shown in Fig. 1.

In regards to load distribution in term of functionalities between system users and moderator under this implementation scenario, the load of user is relatively low because the article evaluation is not required and all system users are assumed to be trussed; they add only bibliographic information of review/survey articles. Of course, there is no moderator load due to moderator is absent in this scenario.

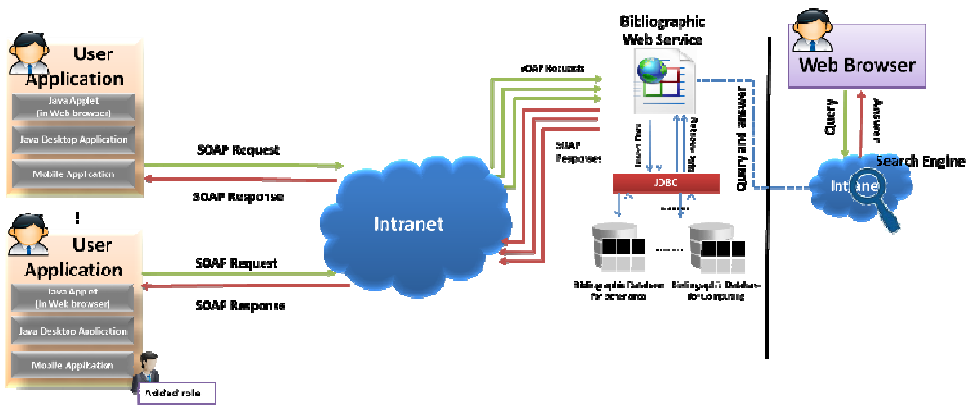

Fig. 1. Architecture of the proposed system (Implementation Scenario 1)

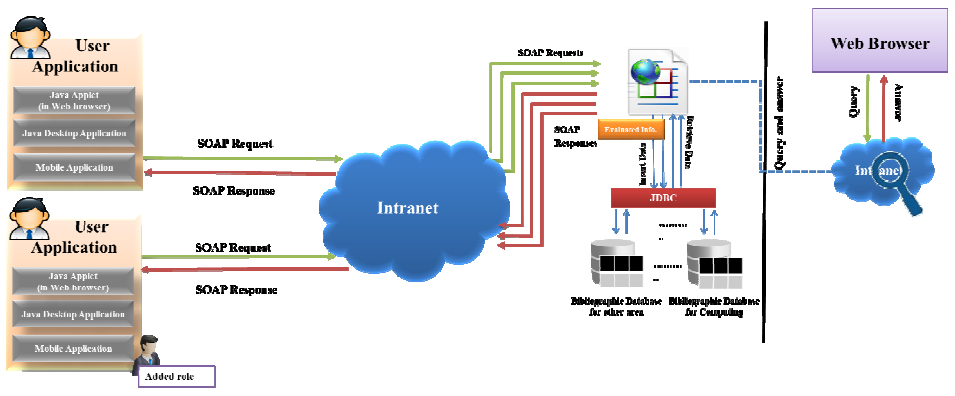

Figure 2: Architecture of the proposed system (Implementation Scenario 2)

***Implementation Scenario 2:***the proposed system will be deployed in private environment to allow users exchanging and retrieving bibliographic information of survey and review articles. The users of such system may be untrusted in term of they may publish bibliographic information of non-survey/review articles. Along with that the moderator is absent in this scenario. Thus, this scenario must take the advantage of social evaluation and feedback, which is not aiming to add opinions about the quality or relevance of content; instead, it aims to take the system user's evaluations and feedbacks by allowing them adding their evaluations and feedbacks about the bibliographic information of un-approved articles. This helps in (i) determining review/survey articles from those pending (not approved) articles that added by users as well as (ii) verifying the correctness of categorization of those articles under the relevant field and sub-field in the science and technology area.

With this scenario, taking user's evaluations and feedbacks for bibliographic information of un-approved articles is made possible. In this scenario, the system user will publish their bibliographic information of articles whether this information for review articles or non-review articles through their applications. This information will be directly open for evaluating by users. The system users will evaluate these articles according to two important things, the first thing is determining the article is review article or not, and the second thing is determining the best classification for this article under relevant field and sub-field. The bibliographic information of review article will be automatically approved and added to system database once the system receiving a certain number of similar evaluation from users. Under this scenario, the system user can show the list of approved review articles that are already evaluated by users as well as the list of un-approved articles that open for evaluations and feedbacks. Fig. 2 shows the architecture of proposed system (scenario 2).

Additionally, the roles of moderator in this scenario is confined only in adding and updating the classification system(s) of science and technology area(s), and it will not be responsible for approving or rejecting the bibliographic information of review articles added by users; which is the role of users. Thus, one of system users aka "Associate user" will take the role of adding and updating the classification system(s) of science and technology area(s) along with its normal roles. The architecture of the proposed system (Scenario 2) is shown in Fig. 2.

In regards to load distribution in term of functionalities between system users and moderator under this implementation scenario, the load of user is relatively high becausethe moderator is absent, the article evaluation is required and it is the only way to make approval to add bibliographic information of review/survey articles. Of course, there is no moderator load due to the absence of the moderator in this scenario.

***Implementation Scenario 3:***the moderator is presence in this scenario with no social evaluation and feedback features. Thus, this scenario is heavily relying on the moderator and does not take advantage of social evaluation and feedback (i.e. user's evaluations and feedbacks) to identify review articles and categorize then under relevant field. It allows system users to publish their bibliographic information of review articles through their applications to administrator/moderator that who will be fully responsible for either approving or rejecting to add such information in database(s) in a whole or in a part. The moderator will ensure that the provided information is for review/survey articles and these articles are under the appropriate field and sub-field prior to take approval decision. The moderator should have enough experience to identify review article from other types of articles and to categorize such article under appropriate field and sub-field. Of course, more than one moderator is a possible way to cope with different science and technology areas, where each moderator is responsible for one science and technology area. The benefit of such scenario is assured that the bibliographic information found in database(s) are valid and accurate. The architecture of the proposed system (Scenario 3) is shown in Fig. 3.

In regards to load distribution in term of functionalities between users and moderator under this scenario, the load of user is relatively low due to the article evaluation is not required as well asthe moderator is present and responsible for all approval/rejection decisions. Of course, moderator load is medium to high due to (1) moderator is responsible for making approval or rejection decisions on all bibliographic information of articles that pushed by system users and (2) the number of users in private environment is not like the number of users in public environment (so less users → less requests).

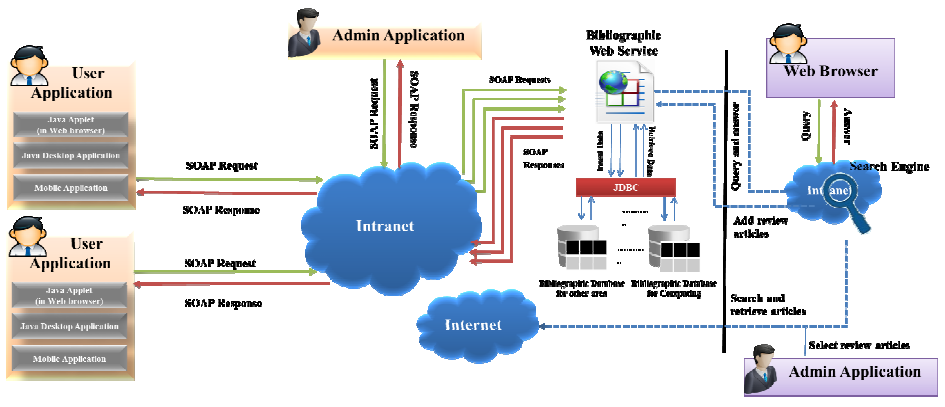

Figure 3: Architecture of the proposed system (Implementation Scenario 3)

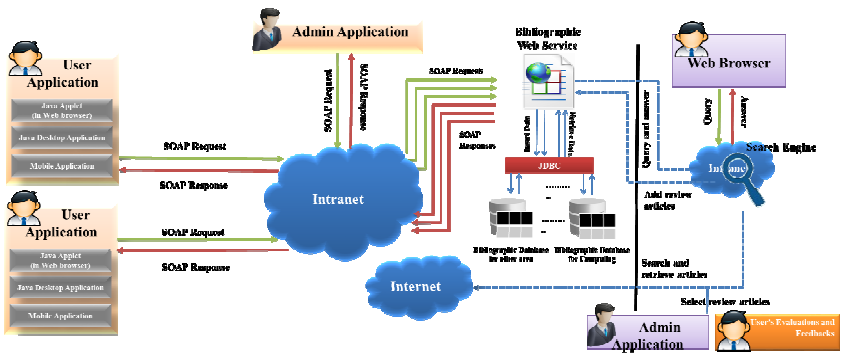

Figure 4: Architecture of the proposed system (Implementation Scenario 4)

***Implementation Scenario 4:***in this scenario, the moderator exists and social evaluation and feedback functionality is supported to allowsystem users adding their evaluations and feedbacks about the bibliographic information of un-approved articles. These evaluations are needed to determine review/survey articles and categorize the bibliographic information of each one of them under appropriate field and sub-field under a specific science and technology area. This scenario is applicable in public environment more than private one.

In this implementation scenario, the users will publish their bibliographic information of review articles through their applications, which may contain bibliographic information of non-review/survey articles. As well, in some cases bibliographic information pushed by users could be difficult for moderator to take decision to approve or reject this information. Thus, the moderator will make this bibliographic information available for evaluation by system users. The system users will evaluate each articleto identify whether it is review/survey article or not and select the best classification for this article under relevant field and sub-field. Of course, the system user will show the approved review articles list and un-approved articles list that are waiting for evaluations in order to be accepted or rejecting according to these evaluation. Moreover, the comment feature may be available for system user as future feature to make comments on articles to correct any mistakes made regarding bibliographic information of review articles. Fig. 2 shows the architecture of proposed system (Scenario 4).

In regards to load distribution in term of functionalities between system users and moderator under this implementation scenario, the load of user is relatively medium due to the article evaluation is required and he/she is responsible to evaluate the un-approved articles that open for evaluations and feedbacks. The moderator load is low to medium due to (1) moderator is responsible for making approval or rejection decisions on some bibliographic information of articles and making the rest open for evaluation and feedbacksand (2) the number of users in private environment is not like the number of users in public environment (so less users → less requests).

On other hand, the proposed system can be implemented and deployed in public environment. In public environment, the administrator or moderator must moderate the proposed system to ensure that bibliographic information of review articles is only added in the system. The implementation of the proposed system in such environment can be carried-out in two different scenarios. These scenarios are describing in the following paragraphs.

***Implementation Scenario 5:***This scenario is Administrator/Moderator without social evaluation and feedback. It is heavily relying on the moderator and does not take advantage of social evaluation and feedback (i.e. user's evaluations and feedbacks) to identify review articles and categorize them under relevant field. It allows system users to push their bibliographic information of review articles through their applications to administrator/moderator that who will be fully responsible for either approving or rejecting to add such information in database(s) in a whole or in a part. The moderator will ensure that the provided information is for review/survey articles and these articles are under the appropriate field and sub-field prior to take approval decision. The benefit of such scenario is assured that the bibliographic information found in database(s) is valid and accurate. Fig. 5 shows the architecture of proposed system (Scenario 5).

In regards to load distribution in term of functionalities between system users and moderator under this implementation scenario, the load of user is relatively low due to the article evaluation is not required as well as the moderator is present and responsible for all approval/rejection decisions. Of course, the moderator load is relativelyhigh due to the following points (1) moderator is responsible for making approval or rejection decisions on all bibliographic information of articles that pushed by system users and (2) the number of users in public environment is not like the number of users in private environment (so more users → more requests).

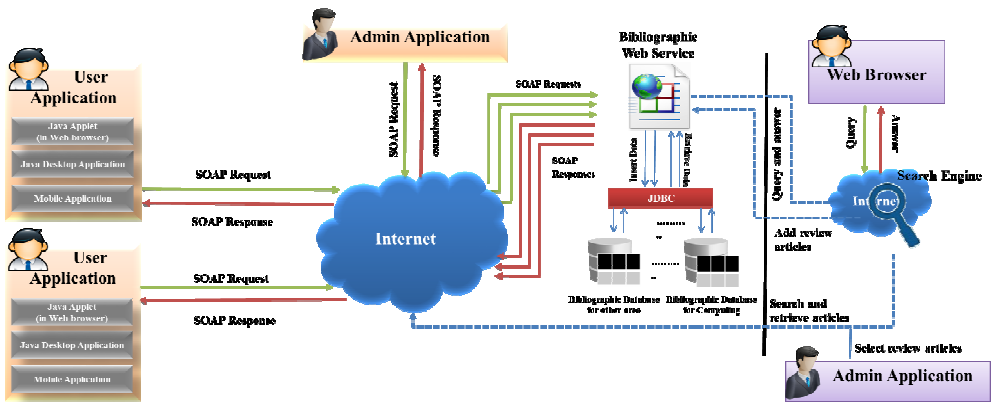

Figure 5: Architecture of the proposed system (Scenario 5)

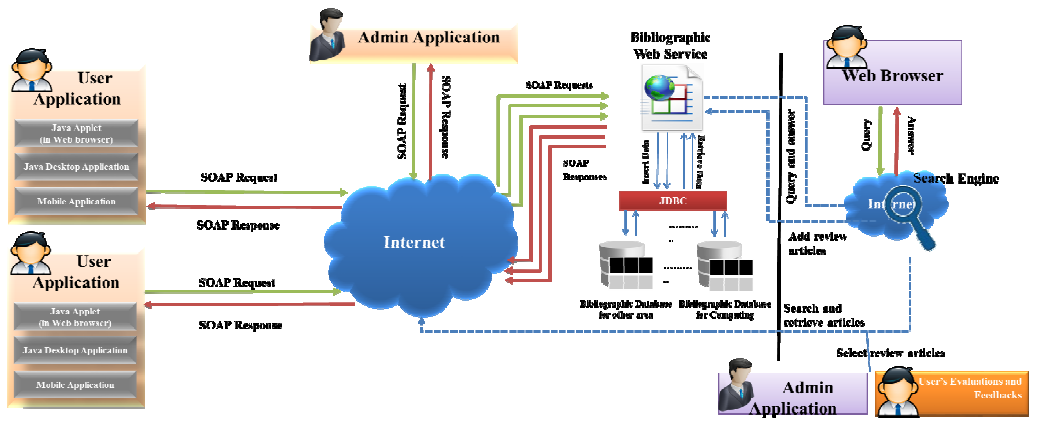

Figure 6: Architecture of the proposed system (Scenario 6)

***Implementation Scenario 6:*** This scenario is Administrator/Moderator with social evaluation and feedback. This scenario benefits from the presence of moderator as well as takes the advantage of social evaluation and feedback, which is not aiming to add opinions about the quality or relevance of content; instead, it aims to take the system user's evaluations and feedbacks by allowing them adding their evaluations and feedbacks about the bibliographic information of un-approved articles. This is helpful in (i) determining review/survey articles from those pending for approval articles that added by users as well as (ii) verifying the correctness of categorization of those articles under the relevant field and sub-field in the science and technology area. With this scenario, taking user's evaluations and feedbacks for bibliographic information of un-approved articles is made possible. In this scenario, the system user will publish their bibliographic information of review articles through their applications. This information may contain bibliographic information for articles that could sometimes be difficult for moderator to take decision to approve or reject this information. In that case, the moderator will make the bibliographic information of articles that are not approved nor rejected available for evaluation by system users. The system users will evaluate these articles according to two important factors, the first one is determining if the article is a review article or not, and the second one is determining the best classification for this article under relevant field and sub-field.

As a conclusion, the system user under this environment can view the list of approved review articles as well as the list of un-approved articles that open for evaluations and feedbacks. In addition, the comment feature may be available for system user as future feature to make comments on articles to correct any mistakes made regarding bibliographic information of review articles. Fig. 6 shows the architecture of proposed system and the future search engine that will be developed later on in a public environment.

In regards to load distribution in term of functionalities between system users and moderator under this implementation scenario, the load of user is relatively medium due to the article evaluation is required and he/she is responsible to evaluate the un-approved articles that open for evaluations and feedbacks. The moderator load is relatively medium due to (1) moderator is responsible for making approval or rejection decisions on some bibliographic information of articles and making the rest open for evaluation and feedbacks (2) the number of users in public environment is not like the number of users in private environment (so more users → more requests).

Beside the proposed system (whether the implementation is carried-out in private or public environment) and for non-system users who do not want to use such system, search engine for scientific review/survey articles will be developed allowing them to search about review articles using their standard Web browsers. Of course, this engine will integrate with Bibliographic Web Service to retrieve appropriate bibliographic information to answer the user's queries. In Scenario 3, 4, 5 and 6, this engine will uses the crawler and parser to retrieve and parse the scientific articles available in different digital libraries and providing them to administrator to determine review articles from them in order to add them under appropriate disciplines or open them for evaluation by users to enrich system databases.

In addition, this engine may use available bibliographic system API(s) such as Mendeley API to retrieve review articles only to add them into appropriate database. Table I compares the different implementation scenarios of the proposed system in private and public environment, where Table II compares the functionalities load between system user and moderator under different implementation scenarios.

Table 1. Comparison between implementations of proposed system in private and public environment

| | **Private environment** | | | | **Public environment** | |
|---|---|---|---|---|---|---|
| | **Imp. Scenario 1** | **Imp. Scenario 2** | **Imp. Scenario 3** | **Imp. Scenario 4** | **Imp. Scenario 5** | **Imp. Scenario 6** |
| **Network** | Intranet | Intranet | Intranet | Intranet | Internet | Internet |
| **Moderator** | Not required | Not required | Required | Required | Required | Required |
| **Article Lists** | Only approved list | Approved list and un-approved (evaluation) list | Only approved list | Approved list and un-approved (evaluation) list | Only approved list | Approved list and un-approved (evaluation) list |
| **Social evaluation and feedback** | Not supported | Supported | Not supported | Supported | Not supported | Supported |
| **Search engine** | Local | Local | Local | Local | Global | Global |
| **Crawler and Integration** | Not supported | Not supported | Supported | Supported | Supported | Supported |

Table 2. Functionalities load between different implementation scenarios

| | **User** | **Moderator** |
|---|---|---|
| **Imp. Scenario 1** | Low load due to no evaluation load and user just use the system | NA |
| **Imp. Scenario 2** | High load | NA |
| **Imp. Scenario 3** | Low load due to no evaluation load and user just use the system | Medium/High load |
| **Imp. Scenario 4** | Medium load | Low/Medium load |
| **Imp. Scenario 5** | Low load due to no evaluation load and user just use the system | High load |
| **Imp. Scenario 6** | Medium load | Medium load |

**Back-end system components**

a) Bibliographic Web Service

Web service is a popular platform-independent standard technology, which is autonomous, self-describing, self-contained, modular application that can be described, published, discovered as well as invoked over Internet [10]. It be able to perform an encapsulated function ranging from a simple request-reply to a full business process [10]. A Web service uses WSDL, a standard description language to expose its interface to the outside world and uses SOAP or Representational State Transfer (REST) to be invoked. However, the bibliographic Web service is a generic as well as simple service that allows users to add, search and retrieve bibliographic information of review articles from different disciplines as well as benefit from offering features. It has a direct access to internal database(s) that contains bibliographic information of survey and review articles for each supported science and technology field, which is queried by system users as well as provides interface for adding and inserting new data. Thus, this service interacts with different databases, where each database is dedicated to specific science field and provides the necessary methods to accomplish its intended function. It is responsible for providing functionalities for system user as well as for system administrator.

For system user, the provided functionalities are: (U1) adding a new bibliographic information of review article, (U2) retrieving a list of review articles in specific sub-field, (U3) retrieving Bibliometrics information of selected sub-field, (U4) rating review paper and retrieve its overall rating score, (U5) retrieving a list of recommended review articles based on user's ratings and (U6) submitting their evaluations and feedbacks for un-approved articles.While, the provided functionalities for system administrator are: (A1) retrieving the list of articles that are pending for decision (approval or rejection), (A2) making the bibliographic information of articles that are not approved nor rejected available for evaluation by system users, (A3) adding, modifying or deleting any sub-field in taxonomy of field and (A4) adding a new science and technology field with its sub-fields using its classification system.Of course, the provided functionalities are differencing between private and public deployment. Table III shows the offered functionalities by Bibliographic Service in each environment with different scenarios of implementation.

Table 3. Offered functionalities in private and public environment

| | **Private environment** | | | | **Public environment** | |
|---|---|---|---|---|---|---|
| | **Imp. Scenario 1** | **Imp. Scenario 2** | **Imp. Scenario 3** | **Imp. Scenario 4** | **Imp. Scenario 5** | **Imp. Scenario 6** |
| **U1** | Supported | Supported | Supported | Supported | Supported | Supported |
| **U2** | Supported | Supported | Supported | Supported | Supported | Supported |
| **U3** | Supported | Supported | Supported | Supported | Supported | Supported |
| **U4** | Supported | Supported | Supported | Supported | Supported | Supported |
| **U5** | Supported | Supported | Supported | Supported | Supported | Supported |
| **U6** | Not supported | Supported | Not supported | Supported | Not supported | Supported |
| **A1** | Not supported | All review articles pushed by users are automatically marked as un-approved articles and open for evaluation by users | Supported | Supported | Supported | Supported |
| **A2** | Not supported | | Supported | Supported | Supported | Supported |
| **A3** | Supported (additional role for Associate user) | Supported (additional role for Associate user) | Supported | Supported | Supported | Supported |
| **A4** | Supported (additional role for Associate user) | Supported (additional role for Associate user) | Supported | Supported | Supported | Supported |

Additionally, to prevent unauthorized access and enforces a proper access, the bibliographic Web service applies a suitable authentication mechanism that meets the requirement of having a valid authentication token to invoke its methods. Obtaining this token requires the correct credentials, which are the username and password.

Furthermore, due to exchanging and transferring data over Internetis naturally unsecure and to avoid storing the cleartext version of user password, the hash value of password is used. This hash value will be generated using SHA1 hashing algorithm in the user side prior sending it to Web service in order to get a valid token. The plain text of user password is not transferring over Internet; it just locally used to generate a hash value. Accordingly, this service is also responsible for providing the following operations:

- Register a new user in the bibliographic database by creating a new profile for his or her.
- Check user credentials are valid before enabling his or her performs operations such as adding, searching or retrieving bibliographic information.
- Modify the profile information such as the password, email and other preferences

.

b) Bibliographic Database

It is a relational database dedicates for storing and querying bibliographic information of survey and review articles in particular science and technology subject area such as computing field, and the data of system users to meet the requirements of the proposed system. For each supported area, the particular bibliographic database will be created to store survey and review articles within this area and related information. The tables of bibliographic database are shown in Table IV.

Table 4. Tables of Biblographic Database

| **Table Name** | **Description** |
|---|---|
| **Classification Table** | stores fields (topics) and sub-fields (sub-topics) of the detailed taxonomy of particular science and technology subject area that found in its classification system to be used to categorize review paper into a sub-topic of relevant topic |
| **Review Article Table** | stores the bibliographic information of review papers |
| **Bibliometrics Table** | stores information of the quantitatively analysis for all sub-fields to explore their impact in the area |
| **Article Rating Table** | stores the calculated overall rating scores for review papers that evaluated by system users according to their evaluation for quality and significant of those papers |
| **Article Rating Detailed Table** | stores the detailed rating information that carried out and assigned by system users for review papers |
| **Article Evaluation and Feedback Table** | stores the values of evaluation and feedback for un-approved articles that submitted by system users. |
| **User Profile Table** | stores the user's base information: first name, last name, username, hash value of the password, email, date of created profile and other preferences |

As we mentioned before, the proposed system will support computing area as proof-of-concept, so the computing bibliographic database will be created with the same structures of tables defined above but particular to computing survey/review articles.

***User Application***

It is a computer application that provides a GUI, which implements the intended user experience in terms of screen layout, screen transitions and screen control elements. It is designed to allow different levels of researchers and students calling methods of Bibliographic Web service through GUI to add their bibliographic information of review articlesin particular supported area as well as retrieving bibliographic information of review articles available in specific sub-field under particular area. The aim of providing a GUI is for defining simple windows, menus and other UI components to facilitate the usage of proposed system. The functionalities supported by this application are shown in Table II.

***Admin Application***

It is a computer application with a GUI that provide specific functions to administrator of the proposed system. These specific functions do not intend to manage system users and does not provide extensive administration functionalities for the proposed system; it just provides the functionalities shown in Table II.

***Web Service Client***

It is responsible for retrieving the WSDL document of Bibliographic Web service and parsing it in order extract the list of methods that are available for remote invocation, generating and sending the SOAP request massage to invoke specific method and receiving the SOAP response message as result of invocation. This client is embedding in user application and administrator application to allow those application calls methods of Web service.

***Web Browser API***

An API uses to create a basic Web browser and embeds it into blank rectangular area in Java Swing application (in our case, it is user application). It allows users to search about relevant information that is needed to add bibliographic information of review article with the same application.

# 4. CLASSIFICATION SYSTEM

It is a classification system (or called classification structure) for particular science and technology subject area that provides detailed taxonomy structure of this area including fields and sub-field. It is a multi-purposes system that can be employed in semantic Web applications, being integrated into the capabilities of search, visual topic displays of the digital library and with any other system that needs a taxonomy structure of particular area for any purpose. This classification for particular area is beneficial for categorizing scientific papers and inclusion each one of them under the appropriate sub-field of relevant field, which leads to maintain and ensure that this system remains up-to-date and relevant. It plays a key role in the development of people search interface and in development of any bibliographic system deals with particular area [11]. As we mentioned before, the proposed system will support computing area as an initial phase to proof-of-concept. Thus, the computing classification system will be defined and explained below.

Computing classification system is a classification system for the computing area provides detailed taxonomy structure of computing topics. It is a multi-purposes system that can be employed in semantic Web applications, being integrated into the capabilities of search, visual topic displays of the digital library and with any other system that needs a taxonomy structure of computing topics for any purpose. It reflects the state of the art of the computing discipline as well as is flexibility to accept any structural change as it evolves in the future. Such system is beneficial for categorizing scientific papers and inclusion each one of them under appropriate topic, which leads to maintain and ensure that this system remains up-to-date and relevant [11].

In the computing area, there are two popular computing classification systems devised by Association for Computing Machinery (ACM) and Institute of Electrical and Electronics Engineers (IEEE). The ACM computing classification system is the classification system for computer science that has twelve main topics, while IEEE computing classification system (IEEE Computer Society - Keywords) is an extended version of the ACM computing classification system and has eleven main topics [11] [12].

In this system, we use a combined version for ACM and IEEE computing classification systems to list the majors of computing and their sub-majors in order to facilitate the process of adding new review articles as well as the process searching about review articles in specific computing area. This version has twelve main topics, which are General Literature, Hardware, Computer Systems Organization, Software Engineering, Theory of Computation, Mathematics of Computing, Security and privacy, Human-centered computing, Applied computing/Computer Applications, Social and professional topics, Networks, Information Technology and Systems, Computing Methodologies and Computing Milieux.

***Bibliometrics Feature***

Bibliometrics can be defined as a set of methods to quantitatively analysis of scientific and technological literature [13]. These methods are most often used in the field of library and information science but they have wide applications in other areas to meet different requirements. In research community, the Bibliometrics can be used to reveal the impact of research fields, the impact of a set of researchers (authors) or the impact of a certain scientific paper. Nowadays, Bibliometrics are actually utilized in quantitative research assessment workouts associated with the output of academic, which is starting to impend practice-based research [14]. In the proposed system, we support a Bibliometrics feature that aims to provide the system user with the following information in fields and their sub-field in particular area: the number of review papers, range of publication years, total citation count for all papers under this sub-field, average rating score and others information. This information helps in determining the popularity and the impact of specific sub-field in particular area compared with other sub-fields in the same area. It also gauges the importance and scientific performance of sub-field to give system user other indications such as the citation count.

***Article Rating Feature***

Rating in general defined in Oxford Dictionary as “a classification or ranking of someone or something based on a comparative assessment of their quality, standard, or performance”. Article rating can be defined as a process for assessing and evaluating the quality and significant of scientific paper. It provides concise information about the quality and significant of paper according to user familiarity level with topic of this paper, and it does not reflect user the actual paper quality and significant. However, Article rating feature provides the ability for researchers to rate any review article based on its quality and significant. Thus, two rating scales are constructed and used to rate review article. The first rating scale is 3-point scale that intended to rate the quality level of article from viewpoint of user; this scale has three categorizes, which are low, medium and high. The second scale is also 3-point scale that intended to determine the familiarity level of the user with the topic of article; this scale has three categorizes, which are low, moderate and expert. These rating scales provide fair rating score for any rated review article after applied normalization. Table IV shows the categorizes of these scales and pre-defined points and values that assigned to each categorize. These rating scales provide fair rating score for any rated review article after normalization.

Table 5. 3-point scale and normalized scale

| **3-point scale** | | **Normalized scale** | |
|---|---|---|---|
| Quality level of article | Point(s) | Familiarity level with article topic | Value |
| Low | 1 | Low | 1 |
| Medium | 2 | Moderate | 2 |
| High | 3 | Expert | 3 |

Therefore, each system user can easily rate any article in the system by using these two scales, he/she just select the article that want to rate then select one option from 3-point-scale to determine the quality level of article and select one option from normalized scale to identify his/her familiarity level with article topic. However, to calculate the overall rating score for the article according to different rating points and values assigned by system users, we use a simple and straightforward method with three phases. In the first phase, we calculate the Normalized Rating Score (NRS) for the article for each user rates this article using:

$$\text{NRS (user, article)} = (\, QPontis(C) * \text{FValue(U)} \,) \,/\, 9 \qquad (1)$$

In Equation (1), C is a selected categorizes by user for article quality-level and U is a selected categorizes by user for his/her familiarity level with article topic. In the second phase, we calculate article Normalized Total Score (NTS) as follow:

$$\text{NTS (article)} = \sum_{for\ each\ user\ rating} \text{NRS (user, article)} \qquad (2)$$

In the third phase, we calculate the overall rating score in percentage for any article by dividing the article NTS from Equation 2 by the number of users that rated the article

$$\text{Score (article)} = \frac{\text{NTS (article)}}{\text{\# of users that rated the article}} * 100\% \qquad (3)$$

Finally, the most important benefit of rating feature is to see the impact of a particular paper/article compared with others papers in the same sub-field level in term of quality and significant according to viewpoints of system users. It also uses to support recommendation feature. However, it is worth remembering that all rating scores in the system are represented the users viewpoints and their research experience to assess scientific papers.

***User's Evaluations and Feedbacks Feature***

This feature aims to take the system user's evaluations and feedbacks by allowing them adding their evaluations and opinions about the bibliographic information of un-approved articles. This helping in (i) determining review/survey articles from those pending (not approved) articles that added by users as well as (ii) verifying the correctness of categorization of those articles under the relevant field and sub-field in the science and technology area. With this scenario, taking user's evaluations and feedbacks for bibliographic information of un-approved articles is made possible.

To achieve this goal, two Nominal scales are constructed and used to evaluate and categorize un-approved review article. The first scale answers the following question: is the bibliographic information of article for review article? This scale has two values, which are review/survey article and not review/survey article. The second scale answers the following question: what is the most relevant field and sub-field for the bibliographic information of article? The values of this scale are the list of fields and their sub-fields in particular science and technology area relevant to this article.

By relying on evaluation and feedbacks that submitted by system users for un-approved articles, the approving or rejecting decision to add bibliographic information of each un-approved article can be made using one of the following options:

- By system administrator/moderator – in this case, the moderator will review the user's evaluations and feedback for each un-approved article and take the final decision whether approving or rejecting to add bibliographic information of such article.

- By system itself – in this case, the system will take the final decision whether approving or rejecting according to pre-defined matching threshold that defined by system administrator. For example, if the system detects ten or more matching between the submitted values of scale one and scale two by different users, the system will automatically make approving decision and adds the bibliographic information of such article in the system database.

***Recommendation Feature***

The recommendation feature aims to recommend survey or review articles for system user by using specific engine named recommendation engine. This engine is the modified version of book recommendation system developed by Noam Sutskever [15]. It recommends articles based on user's ratings using one of the most famous techniques called collaborative filtering. The input of such engine are bibliographic information of review articles, interesting system user and user's ratings, where the output is the list of recommended articles for such user. At the current stage, this recommendation engine is not a comprehensive one; it is only baaed on user's ratings. Further improvement will be taken place at upcoming stages for recommendation engine that will consider and include more factors derived from Bibliometricsinformation.

# 5. FUTUREIMPROVEMENTS

- Comment Feature – it allows system user making comments on the published bibliographic information of articles that already found in the system to correct any mistakes regarding thisinformation. For example, the bibliographic information of non-review article may be added by mistake, so by using this feature, the system user can make a comment on this information to ask moderator to remove the bibliographic information of this article from the system. However, comment feature provides system users the ability for making comment (1) whenever they see any wrong information added to the system, (2) found incorrect classification for review article or (3) found incorrect field or sub- field in the classification system for specific science and technology area.
- Extending Recommendation Feature –the recommendation feature in the proposed system recommends articles based on user's ratings using one of the most famous techniques called collaborative filtering. Other recommendation techniques such as citation analysis, context and social network may be supported by recommendation feature in the later version of the proposed system in order to provide more recommendation options for system users.
- Initiative for Adding Research Type in Published Scientific Paper - As we known, any published scientific paper/article includes mandatory parts such as title, authors, abstract, keywords and others. These parts are required for various purposes, e.g. keywords is required to allow different search engines to find such paper according to its keywords. Currently, the published scientific papers are not included the type of research, which aims to determine the type of article/paperwhether may be review/survey, original, case study or other type. Including such information within the paper is important for reader as well as for computerized system. It allows software or systemto detect the type of paper automatically without human intervention. Thus, we start an initiative thatasking all publishers of scientific papers to add the type of research in each scientific paper before being published in the Internet. This allows the future version of the proposed system to automatically parse the type of research within scientific papersand determine review papers from them to add those papers automatically in the system database without human intervention.
- Private Implementation Scenario(s) of Proposed System in Private Cloud Environment –The demand for rapid deployment and run the proposed system with zero configuration, especially in the current emerging technology, cloud computing, requesting from us to package the proposed system and all needed configurations in .vhdx file. By doing that, the process of deploying and running the private implementation of the proposed system became very easy and convenient with no configuration.Thus, if you have a private cloud and you want to run the private implementation scenario of the proposed system within your environment, the solution is simple by creating a new virtual machine using the packaged .vhdx file. Of course, by using such file, the process of running the proposed system is so simple as well as moving the virtual machine from one host to another is also easily with zero-downtime.

# 6. CONCLUSION

The proposed system is a Bibliographic system for survey and review articles that intends to researchers and student at all levels to add, search and retrieve the bibliographic information of review articles in various science and technology subject areas, which led to build-up a dedicated database for each particular supported area to store its corresponding survey and review articles. It provides two main deployment options, the first option is carried-out in private environment for closed community and the second option carried-out in public environment, i.e. Internet for public community. It supports three rich-features, which are Biblio metrics feature to provide Biblio metrics information about each sub-field in particular area, rating feature that allow system user to rate any review article and recommendation feature to recommend review articles based on user's ratings using collaborative filtering. The additional feature is supporting in case of the proposed system deployed in private environment with Scenario 2 and Scenario 4 or in public environment with Scenario 6, which is social evaluation and feedback (i.e. user's evaluations and feedbacks) feature that helps identifying un-approved articles and categorize then under relevant field. The initial version of the proposed system will support only computing area. Thus, under the collaboration umbrella, this system will be a useful and powerful tool for both researchers and students in computing area to benefit from bibliographic information of computing survey and review articles. However, the proposed system is described from a solely theoretical perspective and the implementationis work in progress. Following the implementation of the proposed system, the search engine will be developed.